\documentclass[prb,aps,amssymb,shownopacs,twocolumn,superscriptaddress,times,10pt]{revtex4-1}
\usepackage{amsmath}
\usepackage{amssymb}
\usepackage{amsthm}
\usepackage{amsfonts}
\usepackage{listings}
\usepackage{enumerate}
\usepackage{latexsym}
\usepackage{float}
\usepackage{psfrag}
\usepackage{bm}
\usepackage[all]{xy}
\usepackage{graphicx}
\usepackage{subfigure}
\usepackage{braket}
\usepackage[pdftex,colorlinks=false]{hyperref}
\usepackage{xcolor}
\usepackage{verbatim}
\usepackage{morefloats}

\usepackage{times}
\usepackage[normalem]{ulem}   
\newcommand{\beq}{\begin{equation}}
\newcommand{\eneq}{\end{equation}}
\newcommand{\bs}[1]{\boldsymbol{#1}}

\def\be{\begin{equation}}
\def\ee{\end{equation}}
\def\ba{\begin{eqnarray}}
\def\ea{\end{eqnarray}}

\def\R{{\rm Re}}
\def\Z{\mathbb{Z}}
\def\C{\mathbb{C}}

\def\Tr{{\rm Tr}}

\def\beq{\begin{equation}}
\def\eeq{\end{equation}}
\def\barray{\begin{eqnarray}}
\def\earray{\end{eqnarray}}

\font\upright=cmu10 scaled\magstep1
\def\stroke{\vrule height8pt width0.4pt depth-0.1pt}

\def\Zmath{\mathbb{Z}}
\def\Qmath{\vcenter{\hbox{\upright\rlap{\rlap{Q}\kern
                   3.8pt\stroke}\phantom{Q}}}}
\def\Nmath{\vcenter{\hbox{\upright\rlap{I}\kern 1.7pt N}}}
\def\Cmath{\vcenter{\hbox{\upright\rlap{\rlap{C}\kern
                   3.8pt\stroke}\phantom{C}}}}
\def\Rmath{\vcenter{\hbox{\upright\rlap{I}\kern 1.7pt R}}}
\def\Z{\ifmmode\Zmath\else$\Zmath$\fi}
\def\Q{\ifmmode\Qmath\else$\Qmath$\fi}
\def\N{\ifmmode\Nmath\else$\Nmath$\fi}
\def\C{\ifmmode\Cmath\else$\Cmath$\fi}
\def\R{\ifmmode\Rmath\else$\Rmath$\fi}

\input{epsf}

\newcounter{defcounter}
\begin{document}

\tolerance 10000

\newcommand{\cbl}[1]{\color{blue} #1 \color{black}}

\newcommand{\vk}{{\bf k}}

\title{Probing many-body localization with neural networks}

\author{
Frank Schindler}
\address{
 Department of Physics, University of Zurich, Winterthurerstrasse 190, 8057 Zurich, Switzerland
}

\author{
Nicolas Regnault}
\address{Laboratoire Pierre Aigrain, D\'epartement de physique de l'ENS, Ecole normale sup\'erieure, PSL Research University, Universit\'e Paris Diderot, Sorbonne Paris Cit\'e, Sorbonne Universit\'es, UPMC Univ. Paris 06, CNRS, 75005 Paris, France}

\author{
Titus Neupert}
\address{
 Department of Physics, University of Zurich, Winterthurerstrasse 190, 8057 Zurich, Switzerland
}

\begin{abstract}
We show that a simple artificial neural network trained on entanglement spectra of individual states of a many-body quantum system
can be used to determine the transition between a many-body localized and a thermalizing regime.
Specifically, we study the Heisenberg spin-1/2~chain in a random external field. We employ a multilayer perceptron with a single hidden layer, which is trained on labeled entanglement spectra pertaining to the fully localized and fully thermal regimes. We then apply this network to classify spectra belonging to states in the transition region.
For training, we use a cost function that contains, in addition to the usual error and regularization parts, a term that favors a confident classification of the transition region states.
The resulting phase diagram is in good agreement with the one obtained by more conventional methods and can be computed for small systems. In particular, the neural network outperforms conventional methods in classifying individual eigenstates pertaining to a single disorder realization. It allows us to map out the structure of these eigenstates across the transition with spatial resolution.
Furthermore, we analyze the network operation using the dreaming technique to show that the neural network correctly learns by itself the power-law structure of the entanglement spectra in the many-body localized regime.
\end{abstract}

\date{\today}

\maketitle


\section{Introduction}

Artificial neural networks are routinely employed for data classification. They are useful when features distinguishing one class of data from another are unknown or unwieldy. A neural network can learn such features from examples, i.e., a set of labeled training data. In physics, the application of neural networks, and machine learning in general, to many-body quantum mechanics is a novel and burgeoning field of research.~\cite{Zdeborova17} Currently, there are three main lines of pursuit: The application of machine learning to the problem of classifying various phases of matter\cite{Broecker2016,Chng16,Frank16,Deng16,Ohtsuki16,Carrasquilla17,vanNieuwenburg17,Hu2017}, accelerating material searches and design\cite{Kusne:2014qv,Ghiringhelli15,Kalinin:2015ty,LiBaker16}, and the quest to encode quantum mechanical states in structures mimicking the setup of a neural network\cite{Torlai16,Carleo602,Deng17}. This work is concerned with the first kind of approach. Most previous studies have considered the identification of phases and phase transitions by training neural networks on a large set of prototype configurations. Here, we instead use entanglement spectra~\cite{LiHaldane08}, which in recent years emerged as a powerful tool to characterize a plethora of physical systems, and have been employed for a neural network based detection of phase transitions in Ref.~\onlinecite{vanNieuwenburg17}.

\begin{figure}[t]
\begin{center}
\includegraphics[width=0.46 \textwidth]{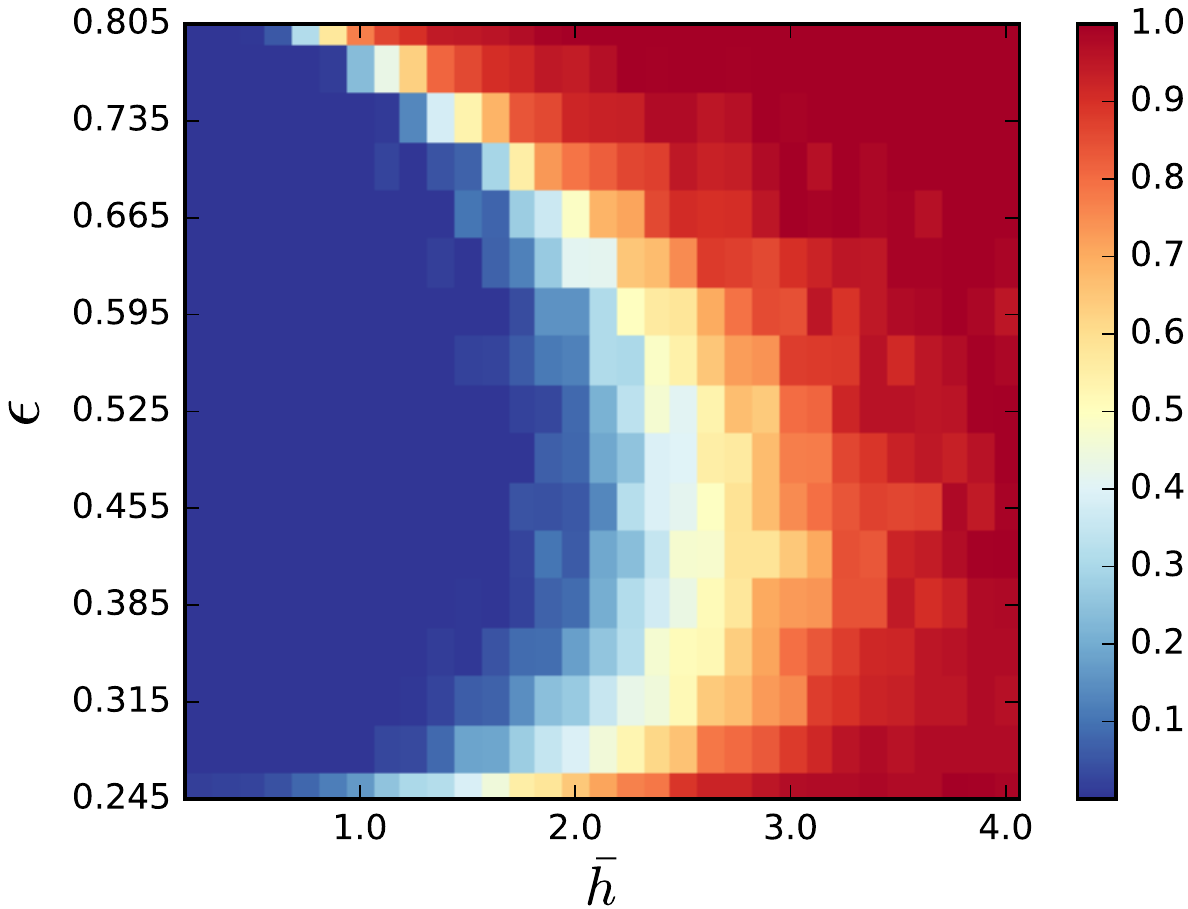}
\caption{Phase diagram of the Heisenberg chain with Hamiltonian~\eqref{eq: HeisenbergH} obtained from the neural network ansatz in Eq.~\eqref{eq: NetworkAction} trained with cost function~\eqref{eq: costfunction} on entanglement spectra obtained from an exact diagonalization on $N=16$ sites. The plot shows the average confidence for the MBL phase over $40$ realizations of disorder as a function of the absolute maximal values of the random magnetic field $\bar{h}$, spaced with $\Delta \bar{h} = 0.125$, and for eigenstates belonging to different rescaled energies $\epsilon = (E-E_{\min})/(E_{\max}-E_{\min})$. Compared to Ref.~\onlinecite{Luitz15} where a similar plot was obtained with better-controlled, yet more sophisticated methods, we have used smaller systems and fewer disorder realizations.}
\label{fig: phasediagram}
\end{center}
\end{figure}

We apply neural network based phase classification to a fundamental question in quantum statistical physics, namely the distinction between systems that obey the eigenstate thermalization hypothesis (ETH) and those violating it. According to the ETH, local observables in a typical many-body eigenstate should take the values that pertain to the observables in a thermal ensemble, with the whole system acting as a heat bath for its subsystems in the thermodynamic limit. A well-studied class of systems that violate the ETH are those exhibiting many-body localization (MBL)~\cite{Anderson58,Fleishman1980,Gornyi05,Basko20061126,Oganesyan14,Huse15,Luitz15,Regnault16}, meaning that partial memory of  initial conditions is preserved for infinite times. Due to this property, which is intimately related to the emergence of an extensive number of integrals of motion\cite{Serbyn2013,Swingle13,Oganesyan14,Ros15}, MBL systems have been envisioned as particularly robust quantum memories.~\cite{Huse13}
Here, we study the Heisenberg chain in a random field as a simple model for MBL. At strong disorder, the model is in the MBL regime, whereas it satisfies the ETH if disorder is weak. Several measures or quantities allow a well-controlled quantitative distinction of thermal and localized regimes. They have been used to study the ETH-MBL transition in finite size numerical simulations, in particular for an extensive analysis of the Heisenberg model in a random field. These characterizing quantities include energy level statistics~\cite{Avishai02,Oganesyan07,Pal2010,Modak14,Agarwal15,Lev2015}, level statistics~\cite{ZhiCheng15,Regnault16} as well as density of states\cite{Serbyn16} analyses of the entanglement spectrum and studies of the distribution of the entanglement entropy over a region of energy eigenstates~\cite{Bauer2013,Pollmann14,Luitz15,Khemani16,Singh16,Monthus16,Yu2016}. Necessarily, these methods rely on a physical understanding of the nature of either regime or of the transition. The neural network based method for identifying the ETH-MBL transition that we present here requires only that the information for distinguishing the ETH from the MBL regime is -- in some form -- contained in the entanglement spectrum. This is useful in particular in situations where the physical characteristics of a phase are not fully understood, as one may certainly argue to be the case for MBL.~\cite{deRoeck16,Prelov16}
Thus, the neural network approach also allows for the possibility of finding ways of characterizing the phase transition beyond established methods, with the network's architecture providing a variational ansatz for a classification criterion.

We use the network to classify the entanglement spectra of \emph{all eigenstates} of the Heisenberg chain, which are obtained by exact diagonalization, in particular at finite energy density (note that Ref.~\onlinecite{vanNieuwenburg17} has characterized the transition using \emph{ground state} properties of the disordered Heisenberg chain via a neural network-based approach of classifying entanglement spectra). For a specific disorder configuration, this allows for instance to trace the evolution of individual ETH states deep in the MBL regime.~\cite{Vosk15, Potter15, Zhang16,Khemani16} We achieve this by considering the spectra from multiple real-space entanglement cuts as input for the neural network. By averaging over disorder realizations, we obtain a phase diagram, Fig.~\ref{fig: phasediagram}, that indicates the location of the ETH-MBL transition as a function of energy density and disorder strength. It is in good agreement with results obtained using conventional methods.~\cite{ZhiCheng15,Luitz15,Serbyn15,Yu2016}

This paper is organized as follows: In Sec.~\ref{sec: NNbinclass}, which may also be read as a short introduction to neural networks, we introduce the general set-up of the network used here, suited for binary classification of data. Subsequently, in Sec.~\ref{sec: MBLEntEnt} we review  the Heisenberg spin chain in a random field, and define the entanglement spectrum. We then discuss the type of input data as well as the network architecture used for classifying entanglement spectra as MBL or ETH in Sec.~\ref{sec: TDandNNArch}. In Sec.~\ref{sec: Results}, we present our results and compare them to existing methods.

\section{Neural networks for binary classification} \label{sec: NNbinclass}

An artificial neural network is an alternating sequence of affine linear maps and nonlinear functions that are successively applied to input data $x$ giving output $y$. Each pair of maps in this sequence is a \emph{layer} of the network. Let the target space of the $\alpha$th layer of the network have dimension $n_{\alpha+1}$, corresponding to $n_{\alpha+1}$ \emph{neurons}. In this work, we focus on binary classification, where we want to learn a map $f(x)$ from the data set $\{x\}$, represented by vectors $x$ of dimension $n_1$, to the discrete target set $\{(0,1),(1,0)\}$. This representation of the target set is somewhat arbitrary--here, we choose \emph{one-hot} vectors, i.e., vectors with a single non-zero element. Their entries are interpreted as the neurons of the output layer.

The network setup described above now implements a trial map $\hat{f}$, which should approximate the unknown map $f$ as good as possible. One important difference is that while the target space of $f$ is discrete, that of $\hat{f}$ is continuous. This allows for smooth convergence of $\hat{f}$ to $f$. To achieve this, we first train the network by adjusting its parameters to gradually improve its performance on a \emph{training set} 
which is labeled, i.e., for which the output of $f$ is known to be either $(0,1)$ or $(1,0)$ for each $x$. We then apply the network to a \emph{testing set} to evaluate how well it generalizes to classify data that it has not seen before. It is essential to avoid overfitting: with a large number $n_2$ of neurons, the network will learn not only the general rules by which the data can be identified as pertaining to the MBL or ETH regimes. Rather, it will also pick up non-universal features, such as noise specific to the training data set that was used. To improve the generalization capability of the network at this stage, we employ \emph{cross-validation}: we first obtain the training and testing sets by randomly subdividing a large set of labeled data into two parts of equal size, and then average the trained network's output (when applying it to previously unlabeled data) over multiple such training runs.

We now describe the full action of the network on the input data. In the first layer, the input vectors $x$ of dimension $n_1$ are mapped to a space of dimension $n_2$ via an affine linear map $x \mapsto V^{(2,1)} \,x + a^{(2)}$, followed by the application of a nonlinear \emph{activation function} $g_2$ (the nonlinearity of which is required in order to be able to approximate arbitrary maps $f$), so that the full action of the first layer may be written as $x\equiv x^{(1)} \mapsto x^{(2)} = g_2(V^{(2,1)} \,x + a^{(2)})$. Here, $V^{(2,1)}$ is a $n_2 \times n_1$ matrix, and matrix-vector multiplication between $V^{(2,1)}$ and $x$ is implied here as well as below. Each entry of the resulting vector $x^{(2)}$ can be interpreted as the output of an individual neuron, of which there are $n_2$ in total. In general, the first layer is followed by further layers, each of which implements the map 
\begin{equation}
x^{(\alpha)} \mapsto x^{(\alpha+1)} = g_{\alpha+1}\left(V^{(\alpha+1,\alpha)} \,x^{(\alpha)} + a^{(\alpha+1)}\right).
\end{equation} 
The elements of the rows in the matrix $V^{(\alpha+1,\alpha)}$ are called the \emph{weights} of the respective neuron, and the corresponding element of the vectors $a^{(\alpha+1)}$ are referred to as its \emph{bias}. All layers but the last one are called \emph{hidden layers}.
If there are $h$ hidden layers, $x^{(h + 2)} = y$ is the two-component output vector. In the networks we use, all vectors, matrices, and numbers are real.

\begin{figure}[t]
\begin{center}
\includegraphics[width=0.45 \textwidth]{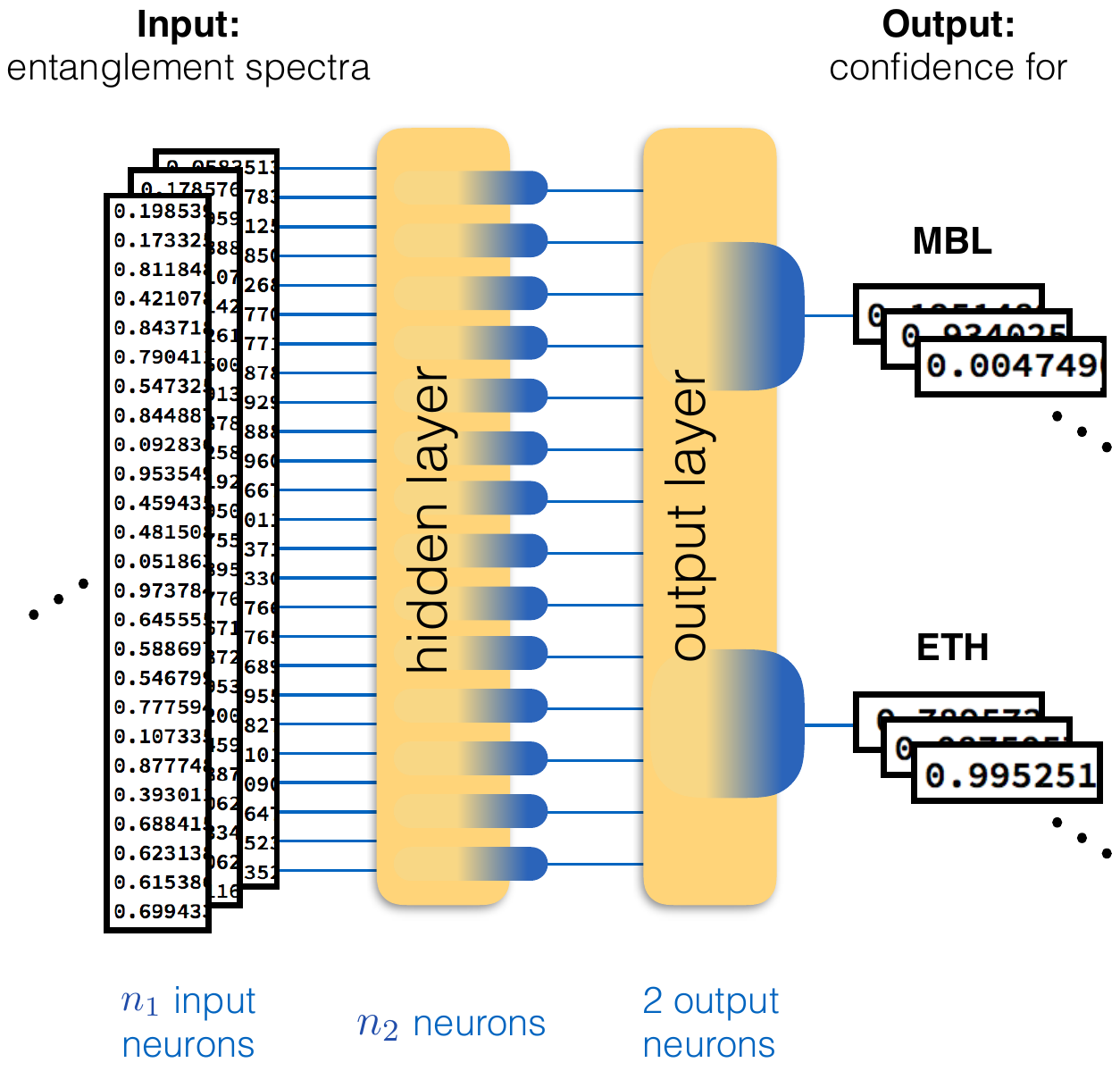}
\caption{Schematic setup of the neural network used to map entanglement spectra to the confidence with which they are classified as either belonging to the ETH or MBL regime. This map, which is explicitly given by Eq.~\eqref{eq: NetworkAction}, can be interpreted as the action of a hidden layer of neurons on the input data, followed by a output layer of two neurons which correspond to the two options of classification. Note that our choice of a $\mathrm{Softmax}$ activation function for the output layer implies that the confidences for ETH and MBL sum up to $1$.}
\label{fig: networkstructure}
\end{center}
\end{figure}

In the following, we will use a network with $h=1$, built from the activation functions $g_2 = \mathrm{ReLU}$ and $g_3 = \mathrm{Softmax}$, defined as
\begin{equation}
\begin{aligned}
\mathrm{ReLU}_i (x) = x_i \, \theta (x_i), \\
\mathrm{Softmax}_i (x) = \frac{e^{-x_i}}{\sum_j e^{-x_j}},
\label{eq: Activation functions}
\end{aligned}
\end{equation}
which are applied component-wise on their vector-valued argument, and the indices $i,j$ run over these components. The projections of the $\mathrm{Softmax}$ output onto the target set vectors sum up to $1$ and can be interpreted as the confidences with which the network allocates the input data to the respective class. For a schematic representation of our network, see Fig.~\ref{fig: networkstructure}.

In order for $\hat{f}$ to approximate $f$, we need to tune the parameters of the network, i.e., the weights and biases, to minimize the discrepancy encoded in an appropriately chosen error functional $\mathrm{Cost}(\hat{f}, f)$. The common choice suited for a $\mathrm{Softmax}$ output layer is the cross entropy
\begin{equation}
\label{eq: CrossEnt}
\mathrm{Cost} (\hat{f}, f) = -\sum_{x \in \{x\}} \sum_{i=1}^2 f_i(x) \log{\hat{f}_i (x)}.
\end{equation}
In training, we then hope to find the global minimum of this functional. One starts from, e.g., randomly initialized weights and biases, which we jointly denote as $X_0$, and then successively applies gradient descent to the weights and biases at step $n$ to obtain those at $n+1$ as
\begin{equation}
\label{eq: updateRule}
X_{n+1} = X_n - \lambda \frac{\partial}{\partial X_n} \mathrm{Cost}, \quad X_n \in \left\{V^{(\alpha+1,\alpha)}_{ij}, a^{(\alpha)}_i\right\}.
\end{equation}
The step size $\lambda$ should neither be too large (otherwise minima are overlooked), nor too small (otherwise convergence is slow and it becomes harder to escape from local minima). A parameter such as $\lambda$, which is not changed during training, but rather determines how we train, is called a hyperparameter. Here, we fix $\lambda$ empirically by requiring optimal minimization of the error on the \emph{training} data. Each such iteration $X_{n} \mapsto X_{n+1}$ of gradient descent is called a \emph{training step}. Since it is too cumbersome to evaluate the error functional for large training sets, we employ stochastic gradient descent; for each iteration, one randomly chooses a relatively small subset of $\{x\}$ as training data. Note that from the point of view of variational calculus, a neural network just corresponds to a shrewd and economic choice of ansatz for minimizing the functional~\eqref{eq: CrossEnt}.

\section{Many-body localization in the Heisenberg chain and entanglement spectrum} \label{sec: MBLEntEnt}

As a toy model for MBL, we study the Heisenberg Hamiltonian in a random field in $z$ direction,
\begin{equation}
\label{eq: HeisenbergH}
H = J \sum_{\mathsf{r}=1}^{N-1} \bs{S}_\mathsf{r} \cdot \bs{S}_{\mathsf{r}+1} + \sum_{\mathsf{r}=1}^{N} h_\mathsf{r} S_{\mathsf{r}}^z
\end{equation}
on an $N$-site chain of spin-1/2 degrees of freedom with open boundary conditions. Here, $\bs{S} = \frac{1}{2} \bs{\sigma}$ acts on the spin on a given site, with $\bs{\sigma}$ the vector of Pauli matrices, and the $h_\mathsf{r}$, $\mathsf{r}=1,\cdots, N$, are static random external fields taken from a uniform distribution in the interval $[-\bar{h},\bar{h}]$. In the following, we will set $J = 1$. The system is integrable for $\bar{h}=0$.
System realizations with $\bar{h}\ll1$ are in a thermalizing (ETH) regime. 
System realizations with $\bar{h}\gg 1$ are in an MBL regime. Both regimes are characterized by different energy level statistics: the ETH regime exhibits level repulsion obeying the Gaussian orthogonal ensemble (GOE) for the Heisenberg Hamiltonian of Eq.~\eqref{eq: HeisenbergH}. On the other hand, the energy spectrum in the MBL regime has Poisson level statistics.

In between the two limits, the behavior of a specific system being either ETH or MBL
depends on the specific disorder realization and the eigenstate that is considered. Averaging over disorder realizations removes these dependencies, but the transition between ETH and MBL regimes may still depend on the energy density at which the system is probed, which amounts to the existence of a many-body mobility edge.
We will assume that at $\bar{h}=0.25$ and $\bar{h}=12.0$ almost all eigenstates belong to the ETH or MBL regime, respectively. 

A characteristic that has been shown to discriminate between ETH and MBL regimes is the entanglement spectrum. It is defined as follows. Consider the reduced density matrix $\rho_A$ of a system in the pure state $|\Psi\rangle$ obtained by subdividing the Hilbert space into two parts, $A$ and $B$, and tracing out the degrees of freedom of $B$,
\begin{equation}
\label{eq: EntSpecDef}
\rho_A = \Tr_B \ket{\Psi}\bra{\Psi} \equiv e^{-H_{\mathrm{e}}}.
\end{equation}
The last equality defines the entanglement Hamiltonian $H_{\mathrm{e}}$.
Here we are interested in a real-space cut separating regions $A$ and $B$ such that all lattice sites $\mathsf{r} \leq N_A$, for some $0 < N_A < N$, are in $A$, and $B$ is the complement of $A$. The spectrum of $H_\mathrm{e}$ is called the entanglement spectrum, and contains information about the nature of $\ket{\Psi}$.

Several possibilities have been explored to determine from the entanglement properties whether a state $\ket{\Psi}$ at finite energy density and fixed disorder shares the character of the MBL or ETH regime. 
(i) The ``Schmidt gap" $\lambda_1(\rho_A)-\lambda_2(\rho_A)$, where $\{\lambda_j (\rho_A); \lambda_j \ge \lambda_{j+1} \}$ denotes the spectrum of $\rho_A$. Being the difference of the two largest eigenvalues of the density matrix, i.e., of the square of the two largest coefficients in the Schmidt decomposition of the system into $A$ and $B$, it is nearly $0$ for mixed $\rho_A$, typical for the ETH regime, and approximates $1$ for almost pure $\rho_A$, characteristic of the MBL phase~\cite{Gray17}.
(ii) ETH states have volume-law entanglement scaling, while MBL states have area-law entanglement scaling. To discriminate between the two in a one-dimensional system, one computes the entanglement entropy $S(N_A)$ as a function of  $N_A$. Extensive scaling of $S(N_A)$ with $N_A$ is expected in the ETH regime, while $S(N_A)$ is constant over different values of $N_A$ in the MBL regime.
(iii) The standard deviation $\sigma_E$ of a sample of entanglement entropies calculated from eigenstates in a range of energies $[E,E+\Delta E]$. Within either phase, $\sigma_E$ is small, while near the transition, where we find both MBL-like and ETH-like states in the energy interval that is probed, $\sigma_E$ is enhanced~\cite{Pollmann14,Singh16,Khemani16,Yu2016}.
(iv) The level spacings in the entanglement spectrum follow distinct statistical distributions in the ETH and MBL regimes. A statistical analysis of the level distributions thus allows to identify the nature of individual eigenstates \cite{ZhiCheng15,Regnault16}.

The power of the neural-network based approach of classifying entanglement spectra as ETH or MBL that we pursue here is that it does not require any a priori knowledge of such criteria. Indeed, the neural network is expected to learn them by itself from the training by examples. In Sec.~\ref{sec: Results}, we compare its performance with (i) and (iii), as well as with the \emph{energy} level statistics.
\section{Training data and network architecture} \label{sec: TDandNNArch}

\begin{figure}[t]
\begin{center}
\includegraphics[width=0.48 \textwidth]{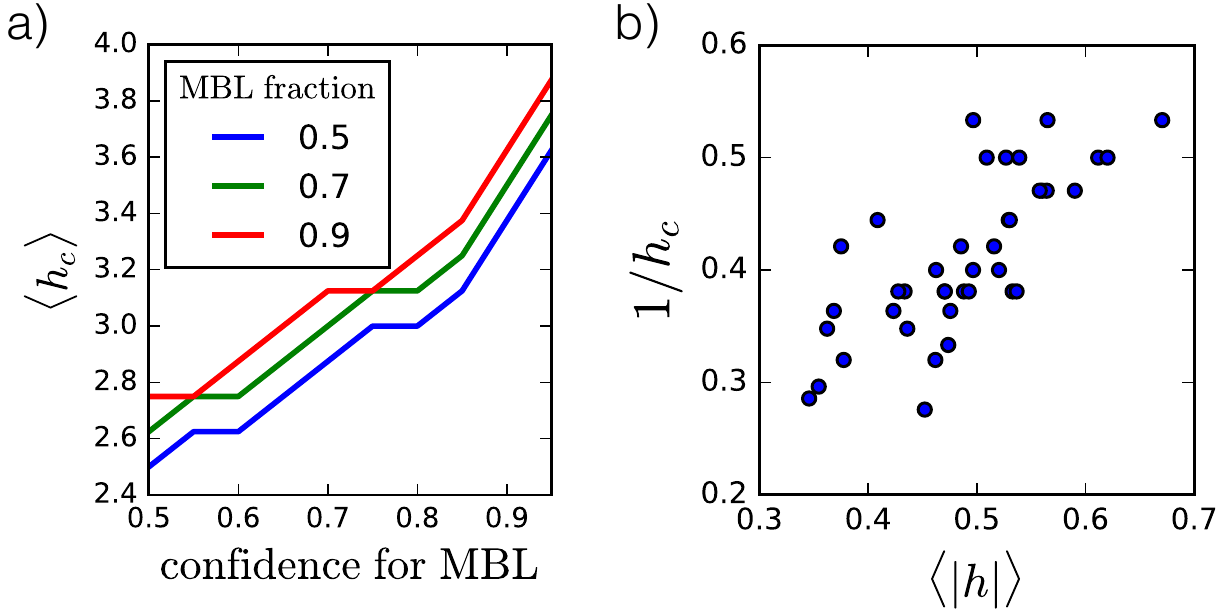}
\caption{
(a) Dependence of the critical value $\bar{h}_{\mathrm{c}}$ on how confident the network is required to be in classifying a given entanglement spectrum as MBL for the average of $40$ disorder realizations of the $N=16$ chain. Here, different lines denote different percentages of MBL spectra that are required in order to classify a given $\bar{h}$ as MBL. For the transition value $\bar{h}_{\mathrm{c}}$, we then take the smallest $\bar{h}$ that is classified as MBL in this way. The plateaus come from the finite $\bar{h}$ resolution $\Delta \bar{h} = 0.125$.
(b)
Correlation of the critical values $\bar{h}_{\mathrm{c}}$ obtained from individual disorder realizations with respective mean disorder strength $\langle|h|\rangle$, averaged over all sites of the $N=16$ chain, for $40$ individual disorder realizations. The correlation coefficient in this case is $\rho = \mathrm{cov}(\bar{h}_{\mathrm{c}},\langle |h|\rangle)/[\sigma(\bar{h}_{\mathrm{c}}) \sigma(\langle |h| \rangle)] \approx 0.76$, with $\mathrm{cov}$ denoting the covariance, and $\sigma$ the standard deviation, respectively.
}
\label{fig: transitiondependence}
\end{center}
\end{figure}

We train with a single-hidden-layer neural network aimed at binary classification of entanglement spectra for eigenstates obtained from the exact diagonalization of Hamiltonian~\eqref{eq: HeisenbergH}.  
For a $N$-site chain, there are $|\{x\}|=2^N$ eigenstates.
Notice, however, that the total spin-projection in $z$ direction measured by the operator $S^z_{\mathrm{tot}} = \sum_{\mathsf{r}=1}^{N} S^z_\mathsf{r}$ commutes with the Hamiltonian~\eqref{eq: HeisenbergH}, corresponding to a global spin rotation symmetry. 
In the following we focus on eigenstates in the $S^z_{\mathrm{tot}}=0$ sector. In the $S^z_{\mathrm{tot}}=0$ subspace, we are thus left with $|\{x\}| = {N \choose  N/2 }$ states, where we only use chains with even $N$ here.

For a cut of size $N_A$ on a $N$-site chain, there are $n_1 = 2^{N_A}$ levels in each entanglement spectrum.
We can further make use of $S^z_{\mathrm{tot},A} = \sum_{\mathsf{r}=1}^{N_A} S^z_\mathsf{r}$ to block-diagonalize the entanglement Hamiltonian.  
From now on, we focus on the largest block, e.g., with $S^z_{\mathrm{tot},A} = 0$ if $N_A$ is even.  In this subspace, the entanglement spectrum has length $n_1 = {N_A \choose \lfloor N_A/2 \rfloor}$, where $\lfloor N_A/2 \rfloor$ is the integer part of $N_A/2$. For training, we additionally leave out the eigenstates at very low and high energies, which are known to deviate substantially from the general trend of the given phase (concretely, we remove the 10\% highest and 10\% lowest energy states). After obtaining the reduced density matrix $\rho_A$ we need to take the logarithm of its eigenvalues to arrive at the entanglement spectrum according to Eq.~\eqref{eq: EntSpecDef}, a procedure, which is prone to numerical errors due to finite machine precision. Hence, we use only the first half (that is, the lower-lying half) of each entanglement spectrum for training, since in both the MBL and ETH regimes, the second half generically consists only of $\rho_A$ eigenvalues which are smaller than $10^{-16}$ and therefore cannot contain any information. Note that the exact size of the part of the entanglement spectrum we train with is irrelevant, we checked that if we instead choose $1/3$ or $2/3$ of it the resulting phase diagram does not change.

\begin{figure*}[t]
\begin{center}
\includegraphics[width= \textwidth]{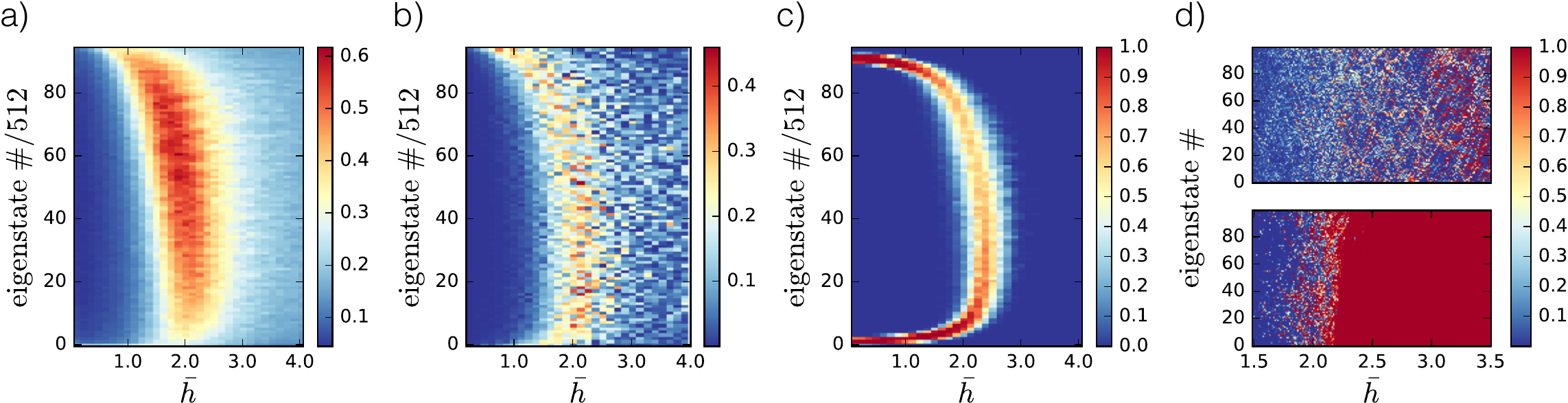}
\caption{Comparison of energy-resolved single-sample ETH to MBL transition indicators. (a) Standard deviation of the von Neumann entanglement entropy of a cut of length $N_A=9$ of the $N=18$ chain over $512$ consecutive eigenstates. (b) Absolute value of the $\bar{h}$ derivative of the Schmidt gap for the same system averaged over $512$ consecutive eigenstates.  (c) Uncertainty in the classification of entanglement spectra by a neural network with one hidden layer, including cross-validation over 50 trainings. States classified as MBL with a confidence larger than $0.1$ but smaller than $0.9$ are assigned a $1$, all others a $0$. The continuous color range comes from averaging over $512$ consecutive eigenstates. (d) Fine structure comparison of the classification of $100$ representative eigenstates taken from the middle of the spectrum of a single sample of the $N=16$ chain (where $N_A=8$). The upper panel shows the value of Schmidt gap, while the lower panel shows the confidence of the neural network for classifying a given state as MBL. Note that this fine-structure analysis cannot be performed with the entanglement entropy standard deviation criterion, as this would require a coarse-graining of eigenstates.}
\label{fig: methodcomparison}
\end{center}
\end{figure*}

We choose the activation functions as in Eq.~\eqref{eq: Activation functions}, so that the full action of the network is
\begin{equation}
\begin{aligned}
\label{eq: NetworkAction}
&\hat{f}(x) = \mathrm{Softmax} [V^{(3,2)} \, \mathrm{ReLU} (V^{(2,1)}\, x + a^{(2)}) + a^{(3)}],
\end{aligned}
\end{equation}
where $V^{(2,1)}$ and $V^{(3,2)}$ are $n_2 \times n_1$ and $2 \times n_2$ weight matrices, respectively, while $a^{(2)}$ and $a^{(3)}$ are the corresponding $n_2$ and two-dimensional bias vectors. Here, $n_1={N_A \choose \lfloor N_A/2 \rfloor}$, and $n_2$ is a free parameter. In fact, we will take $n_2$ to have a relatively large value, of the order of $10^3$. In doing so, the network will become prone to overfit. To avoid overfitting, we employ two strategies (in addition to cross-validation, discussed in Sec~\ref{sec: NNbinclass}).
\begin{enumerate}[(1)]
\item{Dropout regularization: In each training step, we only train with half of the hidden layer neurons, which are randomly chosen each time, effecting the replacement $V^{(2,1)} \rightarrow P V^{(2,1)}$ in Eq.~\eqref{eq: NetworkAction}, where $P$ projects onto a random subset of size $n_2/2$ of the hidden layer units. This prevents successive build-up of neuronal weight configurations adjusted to nonuniversal properties of the training data, and speeds up convergence of the testing set error.}
\pagebreak 
\item{Weight decay: During training, weights that have attained nonzero values at some point, but are no longer actively contributing to the minimization of the error function, should decay to zero in subsequent training steps. This can be achieved by adding a term $-\mu X_n$ on the right-hand side of Eq.~\eqref{eq: updateRule}}, which corresponds to an augmentation of the error functional~\eqref{eq: CrossEnt} by the term $\mu |X|^2$, where $|\cdot|$ denotes the $l_2$-norm of vectors and matrices. Here, we will apply weight decay only to the hidden layer weights $V^{(2,1)}$, since only this preserves a certain reparametrization symmetry of the network.~\cite{Bishop}
\end{enumerate}
With these regularization methods, the number of training steps does not need to be fine-tuned as long as it is large enough. Independent of system size we find that a network with the described architecture classifies samples of testing data, which have no overlap with the training data, very successfully with an accuracy of $\eta = 1$, where $\eta$ is the ratio of correctly identified spectra to all spectra in the testing set. Note that this is the case independent of whether we train and test with entanglement spectra obtained from the same or different disorder realizations.

However, being able to distinguish between the pure ETH and MBL regimes alone, at $\bar{h}=0.25$ and $\bar{h}=12.0$, respectively, is not enough to uniquely determine the classification strategy learned by the neural network. In order to make the predictions for the transition region reliable, we introduce \emph{confidence optimization}: the network should classify entanglement spectra at intermediate $\bar{h}$-values with maximal confidence. Note that this \emph{does not require any prior knowledge} of the phase diagram. To implement this criterion, we add a penalizing term to the error functional which quantifies the lack of confidence at intermediate $\bar{h}$-values. Here, we simply choose the Shannon entropy applied to the network output, since the result of the $\mathrm{Softmax}$ activation function can be interpreted as a probability distribution.

The full error functional used here for training the network to determine the spin chain phase diagram then reads
\begin{equation}
\label{eq: costfunction}
\begin{aligned}
\mathrm{Cost} (\hat{f}, f) = &-\sum_{x \in \text{TD}} \sum_i^2 f_i(x) \ln{\hat{f}_i (x)} \\&- \delta \sum_{x\in \text{TR}} \sum_i^2 \hat{f}_i(x) \ln{\hat{f}_i (x)} + \mu |V|^2,
\end{aligned}
\end{equation}
where $\text{TD}$ stands for training data, i.e., entanglement spectra from $\bar{h}=0.25$ and $\bar{h}=12.0$, while $\text{TR}$ stands for transition region, i.e., entanglement spectra at intermediate disorder strengths $0.25 < \bar{h} < 12.0$. We stress once again that the set for TR is not labeled, meaning we do not make \emph{any} assumption about the nature of the states in the TR. In the last two terms of Eq.~(\ref{eq: costfunction}), $\delta$ and $\mu$ are further hyperparameters controlling the importance of confidence optimization and the strength of weight decay, respectively. We choose suitable values empirically by requiring optimal minimization of the error on the \emph{testing} data. In particular, we observe that as long as both $\mu$ and $\delta$ are chosen to be of order $1$, their exact values do not influence the results significantly. For the following applications, we therefore choose $\delta=\mu=1$, unless otherwise noted. To understand the influence of the respective terms, see Fig.~\ref{fig: costcomparison} in Appendix~\ref{sec: empirical}, where the phase transition regions obtained from networks trained with all possible combinations of $\delta,\mu\in\{0,1\}$ are compared.

We refer the reader to Appendix~\ref{sec: networkdetails)} for further information on the hyperparameters used in Eq.~\eqref{eq: costfunction}, and to Appendix~\ref{sec: empirical} for comparison of results for different system sizes. In all cases, there is no fine-tuning of the network needed. We have checked that changing the hyperparameters slightly from the values we used does not induce noticeable variations in the classification output. 


\section{Results and comparison with conventional methods} \label{sec: Results}

\subsection{Disorder-averaged phase diagram}

With the single-hidden-layer neural network described above, we were able to reproduce the phase diagram of the model given by Eq.~\eqref{eq: HeisenbergH}. Figure~\ref{fig: phasediagram} shows the confidence for the MBL phase averaged over 40 disorder realizations of the $N=16$ chain as a function of the field $\bar{h}$ and the energy density.
A quantitative determination of the critical value of $\bar{h}$ that corresponds to the transition between the ETH and MBL regimes is in part a question of definition. In order to define a critical $\bar{h}_{\mathrm{c}}$, there are two quantities that need to be specified: the threshold for the network confidence, above which a given entanglement spectrum is classified as being in the MBL regime, and the fraction of eigenstates that need to be classified as MBL by lying above this threshold. We show in Fig.~\ref{fig: transitiondependence}~(a)  the resulting dependence of $\bar{h}_{\mathrm{c}}$ on these two quantities for the $N=16$ chain. For example, 
if we consider states above a threshold of $90\%$ confidence as being MBL, and 
require that half of all spectra belonging to a certain value of $\bar{h}$ have to be classified as MBL by this criterion to be at the transition, we obtain a critical value of about $\bar{h}_{\mathrm{c}} = 2.8\pm 0.5$. This agrees with the literature.\cite{ZhiCheng15,Luitz15,Yu2016}

\subsection{Single disorder realization}

To compare the performance with conventional methods in more detail, we consider a specific disorder realization instead of averaging over many. Note, however, that we nevertheless average separately over multiple training runs for cross-validation as explained in chapter~\ref{sec: NNbinclass}. This is also required from the observation of slight deviations in the classification of single eigenstates when the network is trained multiple times, even when this is done with the same input data and training parameters. These deviations can be traced back to the randomized weight and bias initialization we use. See Appendix~\ref{sec: empirical} for a quantitative analysis of this (on average negligibly small) deviation.

To obtain a phase diagram as a function of $\bar{h}$ for a single disorder realization $\{h_\mathsf{r}\}$, we generate $\{h_\mathsf{r}\}$ for $\bar{h}=1$, and then rescale it as $\{\bar{h}h_\mathsf{r}\}$.
Figure~\ref{fig: methodcomparison}~(a) shows the standard deviation of the entanglement entropy over 512 consecutive eigenstates of the $N=18$ chain. The entanglement entropy is expected to be larger in the volume-law entangled ETH regime than in the area law entangled MBL regime. Thus, in the transition region, where some entanglement spectra are MBL-like and some are ETH-like, the entanglement entropy will vary most strongly from one eigenstate to the next.\cite{Pollmann14,Singh16,Khemani16,Yu2016} The maximum in the variance can thus be associated with the transition between the two regimes.

Figure~\ref{fig: methodcomparison}~(c) shows the number of states classified as neither ETH nor MBL for each individual eigenstate of the same system, averaged over 512 consecutive eigenstates. The sharply defined region where this uncertainty is maximal can be interpreted as the neural network's estimate for the transition between ETH and MBL regimes. Note that to obtain this figure, we also performed cross-validation over $50$ training runs.

We compare this result to two established criteria to determine the phase transition, which are well defined for a single disorder realization (and hence also for a fixed system size).
Figure~\ref{fig: methodcomparison}~(a) shows the standard deviation of the von Neumann entanglement entropy. 
Figure~\ref{fig: methodcomparison}~(b) shows the absolute value of the $\bar{h}$ derivative of the Schmidt gap for the same system. As the Schmidt gap is small in the ETH phase and large in the MBL phase, its $\bar{h}$ derivative is expected to attain its maximum value at the transition~\cite{Gray17}.
Comparing Figs.~\ref{fig: methodcomparison}~(a)--(c), we find that the shape and location of the transition agree very well. At the same time, the neural network pins down a much sharper transition than the other approaches.

The power of the neural network based method is even more evident, when the classification of individual eigenstates is considered. Figure~\ref{fig: methodcomparison}~(d) shows that the classification of individual states as belonging to either the ETH or the MBL regime by the neural network is much clearer than by the Schmidt gap criterion.

\begin{figure}[t]
\begin{center}
\includegraphics[width=0.48 \textwidth]{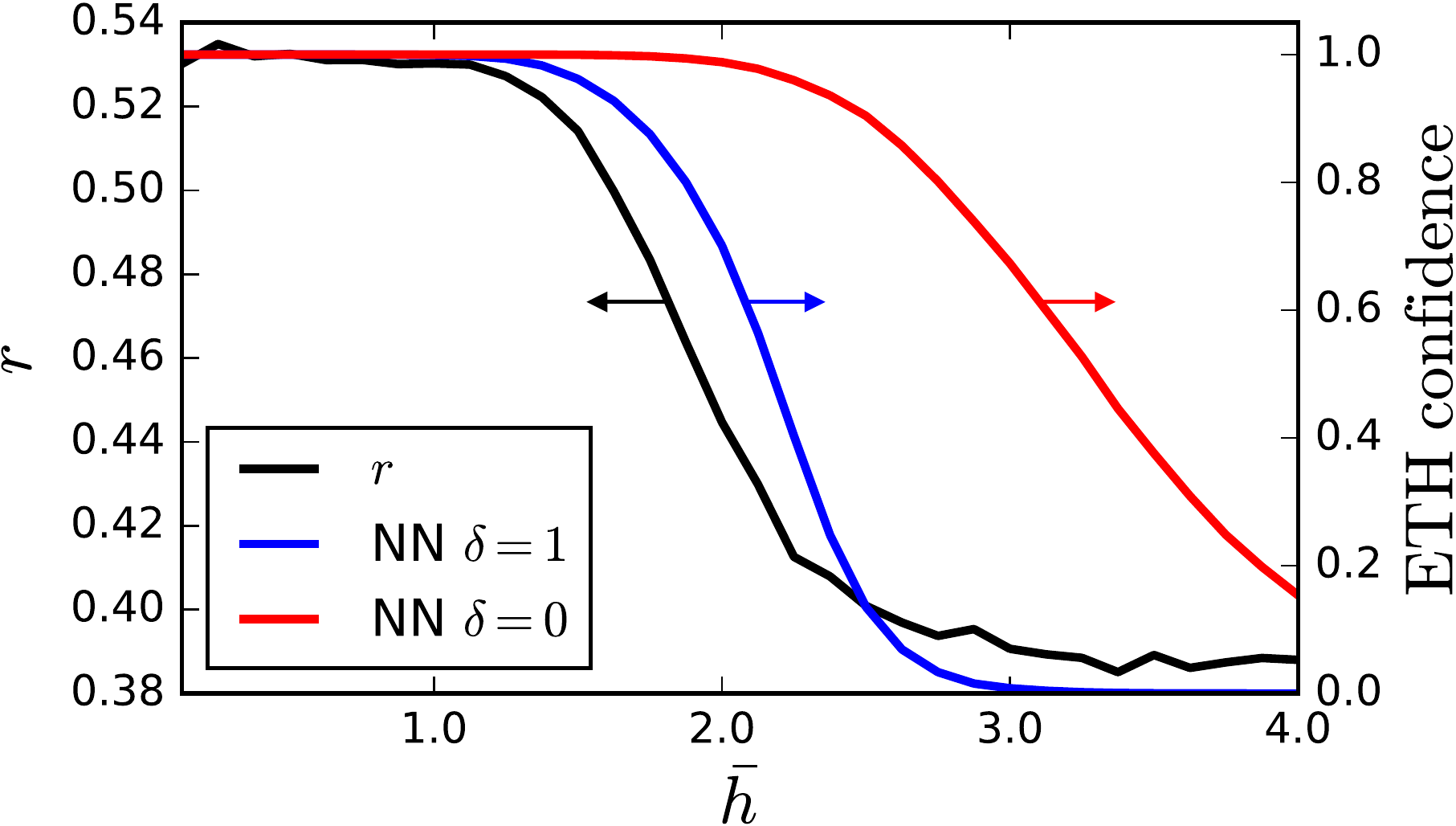}
\caption{Comparison of ETH to MBL transition indicators for a single realization of the disorder of the $N=18$ chain, averaged over energy density. The arrows indicate the vertical axis each curve refers to. The reduced density matrix is built for $N_A=9$. Here, the ratio $r$ of adjacent energy gaps is colored in black. For the GOE describing the ETH regime, we have $r\simeq 0.530$. For the Poisson level statistics characterizing the MBL regime, we have $r\simeq 0.386$. We compare $r$ to the energy-average of the confidence with which entanglement spectra belonging to a given value of $\bar{h}$ are classified as ETH, obtained from two neural network (NN) instances: one with the confidence-enhancing term we added to the network's cost function, corresponding to $\delta=1$ in Eq.~\eqref{eq: costfunction}, the other without it, corresponding to $\delta=0$, colored in blue and red, respectively. The large deviation in the classification of the network trained with $\delta=0$ from the established criterion of energy level statistics underlines the importance of confidence optimization. Note also that the transition is sharper for the neural network based approach with $\delta=1$.}
\label{fig: lvlstatisticscomparison}
\end{center}
\end{figure}

Next, we compare the neural network based transition characterization with the established method of energy level statistics (see Sec.~\ref{sec: MBLEntEnt}). To identify the energy level statistics, we use the average ratio $r$ of adjacent energy gaps.
For an ordered spectrum $\{E_n;E_n \leq E_{n+1}\}$, the ratio of adjacent gaps is defined as

\begin{eqnarray}
r_n&=&\frac{{\rm min}\left(E_n - E_{n-1}, E_{n+1} - E_{n}\right)}{{\rm max}\left(E_n - E_{n-1}, E_{n+1} - E_{n}\right)}
\,.
\label{eq:RatioAdjacentGap}
\end{eqnarray} 
Each energy level distribution leads to a given average ratio $r$ of these $r_n$. The comparison between $r$ and the neural network based transition is shown in Fig.~\ref{fig: lvlstatisticscomparison}. Here it becomes clear that confidence optimization, as represented by an additional term in the cost function [corresponding to $\delta=1$ in Eq.~\eqref{eq: costfunction}], which is designed to favor networks which confidently classify transition region states, makes physical sense and is essential to make contact with established methods.
 
We note that by refining and combining conventional methods\cite{ZhiCheng15,Luitz15,Serbyn15,Yu2016} the same or better classification success can be achieved. However, here we want to point out that there is no equally simple method, assuming as little prior knowledge, that performs equally well as the machine learning based approach. This approach is very basic: we use a single hidden layer, typical neural activation functions, and apply standard regularization techniques. The only nontrivial input we added to this standard setup is cost optimization, which is effected by a non-zero $\delta$ in Eq.~\eqref{eq: costfunction}.

We observe that the location of the transition varies substantially with different disorder realizations sampled from the same distribution characterized by some fixed $\bar{h}$. The main reason for this is that even with fixed $\bar{h}$ per site of the chain, the mean strength of disorder can still fluctuate. As can be seen in Fig.~\ref{fig: transitiondependence}~(b), the disorder strength averaged over all sites $\langle |h|\rangle = \frac{1}{N}\sum_{\mathsf{r}=1}^{N} |h_\mathsf{r}|$ is directly correlated with the inverse of the field corresponding to the transition, $1/\bar{h}_{\mathrm{c}}$.
We thus identify it as a key ingredient for the dependence of the transition on the disorder realization, even though other properties of the disorder realization than the average absolute field value may also be correlated with the location of the transition.

\subsection{Spatial structure of individual eigenstates}

The neural network classification of individual eigenstates based on the entanglement spectrum further allows to analyze the \emph{local} structure of these states, by varying the location of the entanglement cut. We compute for a fixed disorder realization the entanglement spectra of each eigenstate for seven consecutive cuts in the middle of the $N=18$ chain as a function of $\bar{h}$. All of these entanglement spectra are subsequently classified using the neural network. We do not perform cross-validation as we will only include confidently classified spectra in the subsequent analysis. Due to the variation of the length of the entanglement spectra with the cut location, a new training of the network is necessary for each cut.

\begin{figure}[t]
\begin{center}
\includegraphics[width=0.48 \textwidth]{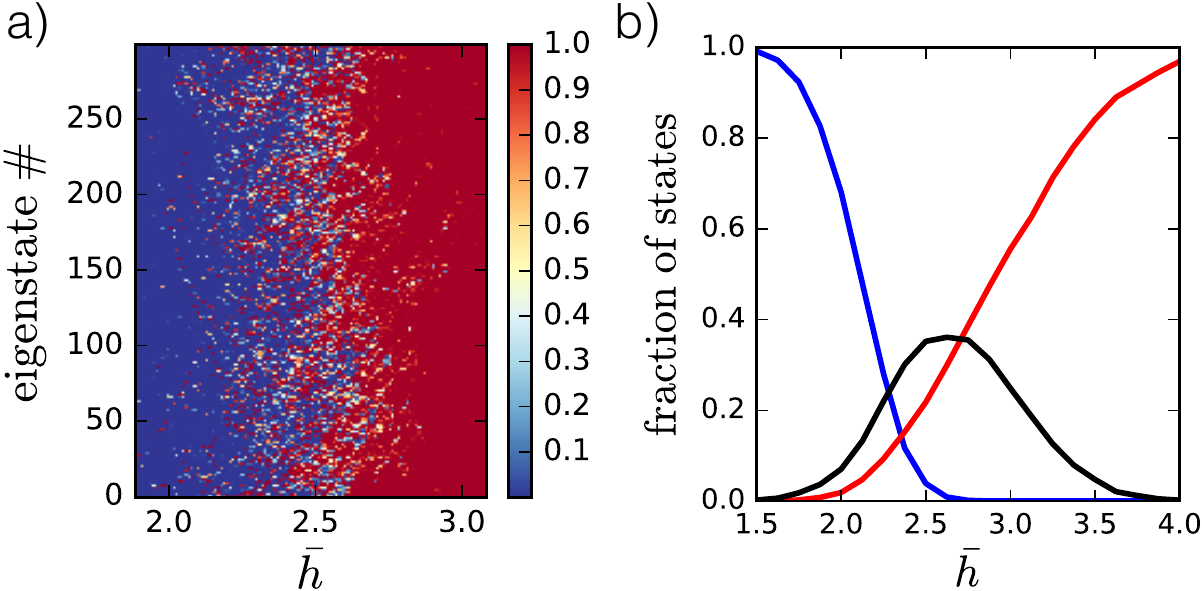}
\caption{Dependence of the classification of eigenstates for a single disorder realization of the $N=18$ chain on the location of the entanglement cut. (a)
Close-up of the $\bar{h}$-dependent classification of $200$ eigenstates taken from the middle of the spectrum for entanglement cuts ranging from $N_A=6$ up to $N_A=12$. For each value of $\bar{h}$, a subplot is shown whose horizontal axis corresponds to these seven cuts.
Blue (red) states were classified with confidence $>90\%$ as ETH (MBL). The remaining states are marked in white.
(b)
Survey of $10\,000$ eigenstates from the middle of the spectrum which have been classified separately for all seven cuts.
The blue (red) curve is the fraction of eigenstates that were classified with confidence $>90\%$ as ETH (MBL) for all cuts. 
The black curve is the fraction of states that are classified with confidence $>90\%$ as ETH for at least one cut and with the same confidence as MBL for at least one cut, i.e., states which are spatially inhomogeneous in their character.  
}
\label{fig: grassplot}
\end{center}
\end{figure}

Figure~\ref{fig: grassplot}~(a) shows the entanglement cut-resolved classification results for 200 consecutive eigenstates as a function of $\bar{h}$, by varying the cut for the entanglement spectrum to lie on seven consecutive bonds. 
The resulting locally resolved classification of eigenstates displays a remarkable asymmetry between approaching the transition from the ETH or from the MBL side; first, we consider the few states that are MBL classified for some entanglement cut deep in the ETH regime. We observe that a substantial fraction of them was classified as MBL consistently across all seven cuts. This is in sharp contrast with the few states that are ETH classified for some entanglement cut deep in the MBL regime. Only a tiny fraction of these states is classified as ETH across all seven cuts simultaneously. 
Our results point to an asymmetry in the evolution of the local structure of eigenstates when the transition is approached from the MBL or ETH side.


Using this capability of resolving the state character locally, we can thus confirm the following hypothesis about the evolution of the local structure of the quantum states across the transition~\cite{Vosk15, Potter15, Zhang16,Khemani16}: 
As the transition is approached from the MBL side, ETH-like regions emerge in rare places where the random field variations are small. These are the first places where quantum fluctuations will dominate over the classical ones induced by the random field. As the transition is approached, these regions grow in size and will eventually be large enough to serve as a bath for the entire system. At this point, the state becomes entirely ETH-like. 
We make these observations quantitative in Fig.~\ref{fig: grassplot}~(b), where the fraction of pure ETH states, pure MBL states, and of states with mixed character is plotted as a function of $\bar{h}$.

We remark that for the above analysis to be valid, it is crucial that we are considering spin chains with open boundary conditions. Only in this case does an entanglement cut separating the chain in two subsystems reveal purely local information. In contrast, if the chain was studied with periodic boundary conditions, a bipartitioning requires two cuts and the entanglement spectrum would convolute information about the structure near both cuts at distant locations.

\subsection{What the network has learned: dreaming}

Taken collectively, the above results on the disorder-averaged phase diagram as well as about the transition for individual disorder realizations provide strong evidence that the neural network performs the classification operation for which it was designed.
To gain further insights into the network operation and to rule out that any pathological behavior emerges, we analyze it with a method called \emph{creation by refinement}\cite{deepDream88}, which has recently gained widespread attention under the term \emph{dreaming}\cite{deepDream15} in the application to image recognition. The method is, however, more general than this, and here we use it on entanglement spectra. For that, the trained network is presented with randomly initialized input data $x$ for which it provides an output $y$. Subsequently, the input data is modified until the output $y$ equals a desired entry from the target set $\{(1,0),(0,1)\}$, for example $(0,1)$ for being classified as MBL. When this procedure is repeated many times, the obtained collection of input data $x$ reveals which properties of the training data have been learned by the network as being characteristic for an MBL entanglement spectrum.

We perform the dreaming algorithm using gradient descent to optimize the input data for a desired network output, using the cross-entropy between the desired and the actual output as a cost function. Instead of using entirely random input data to initialize the dreaming, we randomly choose entanglement spectra that have not been classified confidently as either MBL or ETH by the network (more precisely, we choose spectra for which the confidence of ETH and MBL is between 40\% and 60\%). These input spectra are then optimized toward either ETH or MBL. To obtain comparable and physical entanglement spectra, we further enforce the constraints that (i) the eigenvalues of the reduced density matrix be nonnegative, and (ii) the absolute value
of their logarithm be ordered from smallest to largest. Enforcing these constraints is necessary, because we cannot expect that the network has `learned' any of them, and not restrictive, as any physical entanglement spectrum can be represented this way. Note that we do not require that the eigenvalues of the reduced density matrix sum up to $1$, since this property of the training data was lost when restricting to the first half of the entanglement spectrum as described at the beginning of Sec.~\ref{sec: TDandNNArch}.

\begin{figure}[t]
\begin{center}
\includegraphics[width=0.48 \textwidth]{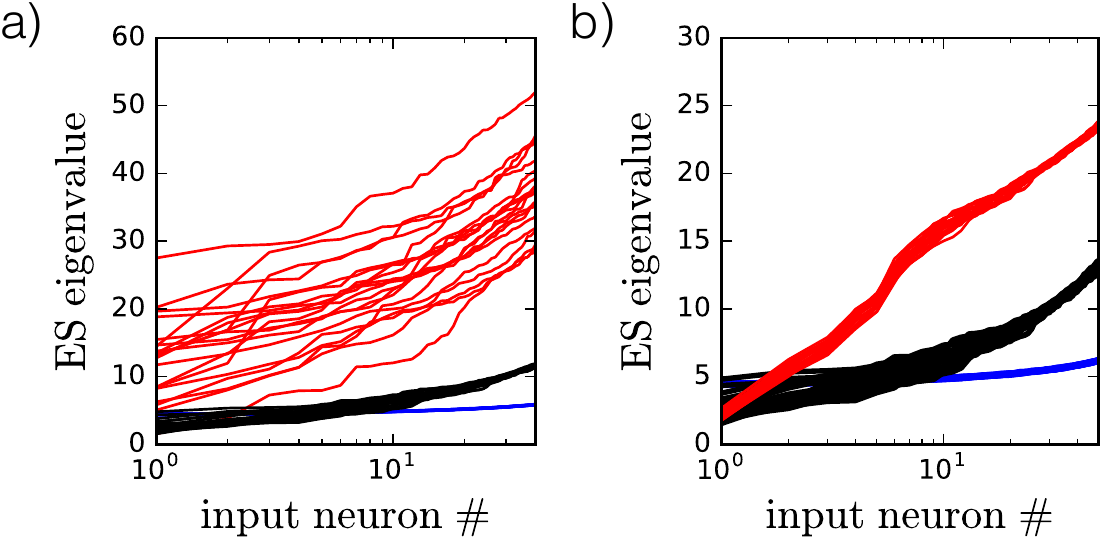}
\caption{Dreaming of entanglement spectra for a single disorder realization of the $N=18$ chain (with $N_A=9$). An entry on the horizontal axis denotes the index of the respective entanglement spectrum eigenvalue. (a) 140 typical entanglement spectra of the pure ETH (blue), MBL (red), and transition (black) regime. The trained network assigns a confidence vector $(1,0)$ to the ETH spectra, and $(0,1)$ to the MBL spectra. The transition regime entanglement spectra have been chosen such that the network yields an output lying between $(0.4,0.6)$ and $(0.6,0.4)$ on them. (b) Result after $200\,000$ steps of the dreaming gradient descent, described in the main text, when applied on 20 of the same transition region entanglement spectra. Note the difference in scale, which indicates that the neural network learns relative rather than absolute features of the training data. The network correctly picks up the power-law structure of the entanglement spectrum in the MBL phase, as can be deduced from its nearly linear slope in this logarithmic plot.
}
\label{fig: deepdream}
\end{center}
\end{figure}

To keep the dreaming algorithm from merely capturing structure specific to one training instance of the network, we average over $40$ distinct training instances. For each dreaming, we use $2 \times 10^5$ steps with a gradient descent step size of $10^{-4}$. The results are shown in Fig.~\ref{fig: deepdream}~(b) where a comparison with true entanglement spectra from deep in the MBL and deep in the ETH phase is made. We observe from the resemblance between Figs.~\ref{fig: deepdream}~(a) and~(b) that the neural network has indeed learned the relative characteristic shapes of typical ETH and MBL entanglement spectra, in particular the power-law nature of the entanglement spectra in the MBL phase,\cite{Serbyn16} but without becoming sensitive to their exact absolute magnitudes. This can be understood by noting that it was only trained to distinguish one phase from the other, and not to individually characterize the respective phases on their own.


\section{Summary} \label{sec: Summary}
We trained an artificial neural network with a single hidden layer on entanglement spectra of the disordered Heisenberg chain to identify the ETH and MBL regimes of this model, and subsequently applied the network to entanglement spectra belonging to states that lie between these two regimes. Even though the network was not trained with entanglement spectra near the phase transition, the phase diagram obtained is in good agreement with previous studies. By adapting the dreaming technique, we were able to show that the network correctly learns the power-law entanglement spectrum of the MBL phase. Our method is uncontrolled, and is therefore less qualified to deduce a quantitative value for the critical disorder strength, for example. However, it has the advantage of being simpler than conventional methods, and in addition provides the cleanest characterization of the transition in the case of single eigenstates of a single disorder realization.

\begin{figure*}[t]
\begin{center}
\includegraphics[width= \textwidth]{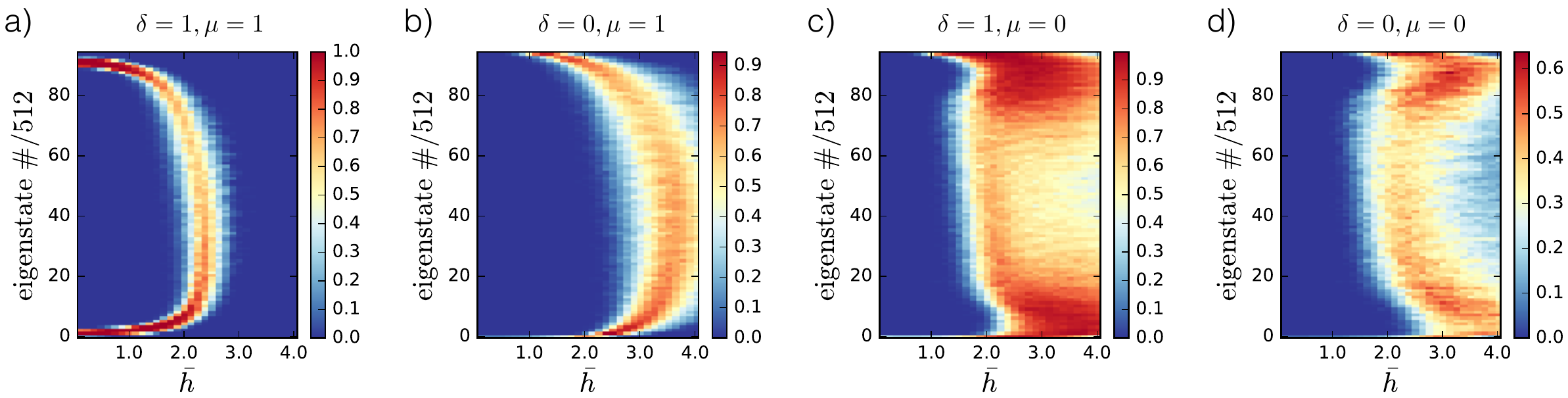}
\caption{
Comparison of transition regions obtained from neural networks that were trained with different cost functions, corresponding to a different choice of the parameters $\delta$ and $\mu$ in Eq.~\eqref{eq: costfunction}, for a single sample of the $N=18$ chain with $N_A = 9$. Here we show the classification uncertainty as defined below Fig.~\ref{fig: methodcomparison}~(c), i.e., including cross-validation over 50 trainings,  for different cost functions. (a) The standard classification result with both regularization and confidence-enhancing terms in the cost function. (b) Without confidence optimization, the transition region is detected at qualitatively larger values of $\bar{h}$. From this observation alone we cannot decide which of the two networks corresponding to $\delta \in \{0,1\}$ should be preferred, since both distinguish the pure ETH and MBL regions with $\eta=1$. However, Fig.~\ref{fig: lvlstatisticscomparison} clearly shows that only the cost function \emph{with} $\delta=1$ faithfully recovers the transition obtained from a level statistics analysis. In addition, as discussed in Appendix~\ref{sec: otherinputdata}, when training with different kinds of input data the transition region stays unchanged only when $\delta=1$, while for $\delta=0$ the network classification becomes inconsistent. (c) and (d) Without proper regularization, the network performance becomes pathological.}
\label{fig: costcomparison}
\end{center}
\end{figure*}

The structure and operation of the network that we presented, including features such as confidence optimization and dreaming, is broadly applicable to classify other phases of matter based on their entanglement spectrum and likely also other correlation functions as input data. We conclude that artificial neural networks constitute an efficient and unbiased tool for classifying phases of matter from numerical data.

\begin{acknowledgments}
We thank Markus M\"uller, Maksym Serbyn, Nils Eckstein, and Ole Richter for useful discussions, as well as Rahul Nandkishore and Scott D.\ Geraedts for previous collaboration on related subjects. In addition we thank Markus M\"uller for a critical reading of the manuscript. F.S. acknowledges support by the Swiss National Science Foundation under grant number 200021-169061.
  \end{acknowledgments}


\appendix

\section{Training with other kinds of input data}
\label{sec: otherinputdata}
In the main text, we have focused on the spectrum of $H_\mathrm{e}$, defined by Eq.~\eqref{eq: EntSpecDef}, as input data for our neural network. In this appendix, we report our observations when training instead with (a) the spectrum of the reduced density matrix $\rho_A$, and (b) the differences of the $H_\mathrm{e}$ eigenvalues. The motivation for input data of type (a) is that diagonalizing $\rho_A$, instead of its logarithm, requires less preprocessing, while the motivation for (b) is that the differences of the $H_\mathrm{e}$ eigenvalues have been shown~\cite{ZhiCheng15,Regnault16} to be statistically distributed in a unique fashion depending on whether the regime is MBL or ETH.

\begin{figure}[b]
\begin{center}
\includegraphics[width=0.45 \textwidth]{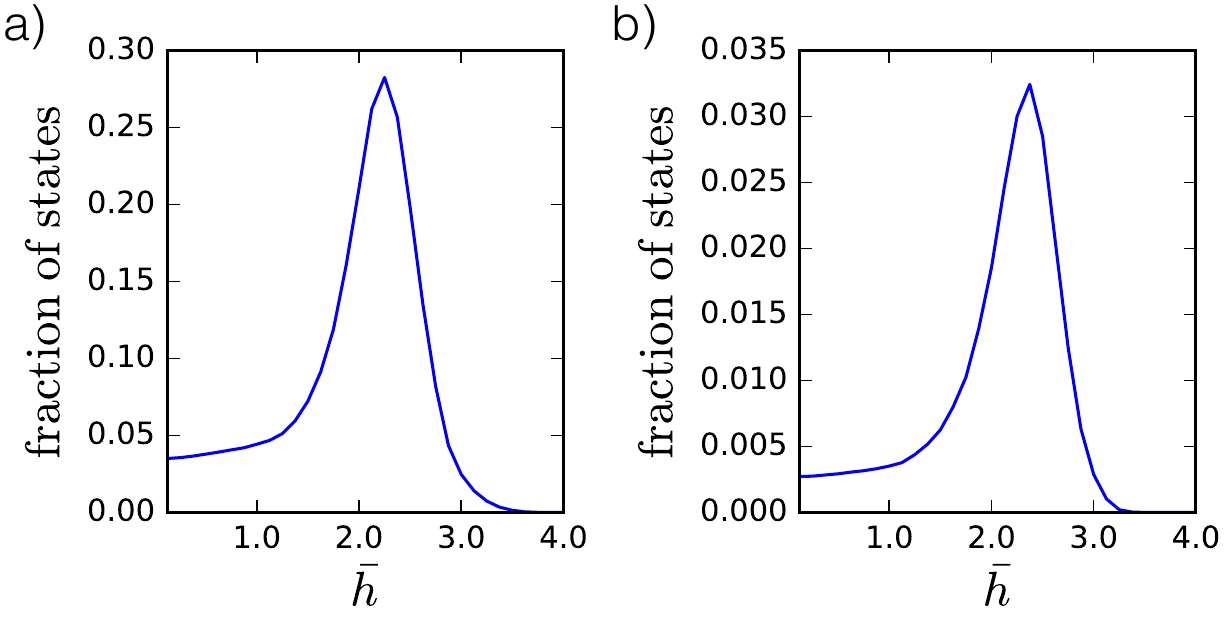}
\caption{
Dependence of the classification of states on different training runs. For each plot, the entanglement spectra for a system with $N=18$ sites with a single disorder realization were trained with 50 times, each time with randomly initialized network parameters and a random choice of training data. The resulting classifications of the individual eigenstates for each $\bar{h}$ are then compared pairwise and the comparison is averaged over all pairings of trainings. We consider three categories of states: (i) classified with confidence $>90\%$ as MBL, (ii) classified with confidence $>90\%$ as ETH, (iii) others, not confidently classified.
(a)
Fraction of states that switch between any of the three categories (i)--(iii) to any other between two trainings.
(b)
Fraction of states that switch from category (i) to category (ii) (i.e., from confidently MBL to confidently ETH) or vice versa between two trainings.
Note that the value of $\bar{h}$ at which these fractions assume their maximal value provides one method to determine the critical transition location $\bar{h}_{\mathrm{c}}$ for this particular disorder realization.}
\label{fig: training dependence}
\end{center}
\end{figure}

\begin{figure*}[t]
\begin{center}
\includegraphics[width= \textwidth]{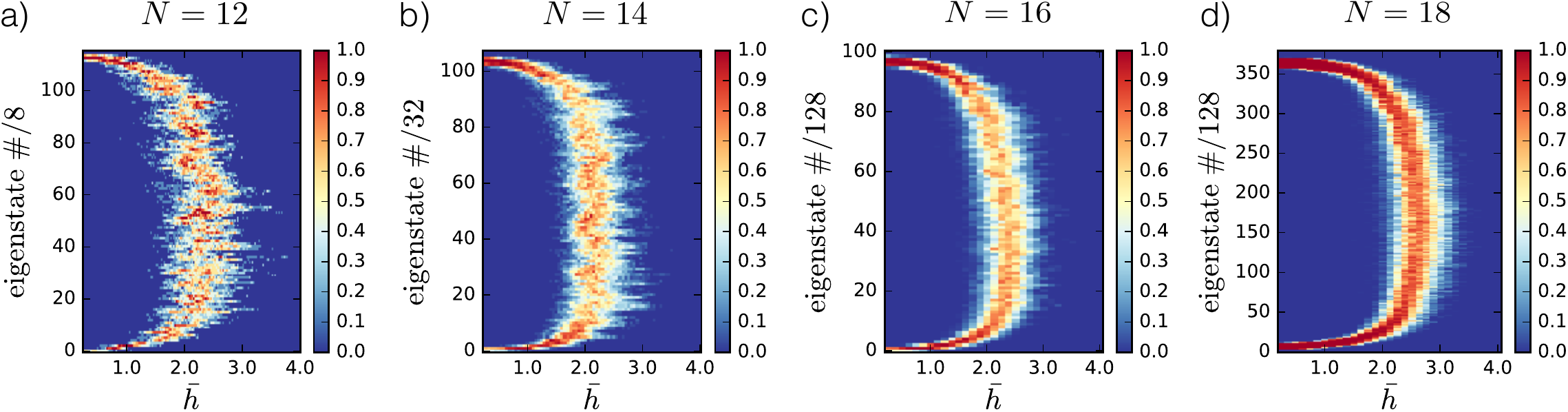}
\caption{
Finite-size dependence of the transition region for different single disorder realizations of the $N = 12,14,16,18$ chains (note that we have on purpose used a different disorder realization for $N=18$ than the one used in Fig.~\ref{fig: methodcomparison} to show that the overall shape of the transition region is sample independent). 
Each plot shows the transition region as defined below Fig.~\ref{fig: methodcomparison}~(c) for a network trained on entanglement spectra of a cut with $N_A = N/2$. Note that the $\bar{h}$ step size for each plot differs, and is given by $\Delta \bar{h} = 0.01, 0.05, 0.125, 0.125$, respectively.
For each $N$, we have averaged the result over an increasing number of eigenstates in order to arrive at a resolution suited for the human eye in a controllable way.
Only for $N=16$ and $N=18$, the data are averaged over the same number of eigenstates, in order to demonstrate that the apparent improvement of the sharpness of the transition is not merely a result of a larger number of states that have been averaged over.
Note also that each of panels (a)--(d) necessarily represent different disorder realizations and hence the value of $\bar{h}$ at which the transition occurs should not be compared between them. 
}
\label{fig: finitesize}
\end{center}
\end{figure*}

In the case of (a), we again are able to train the same network as described in the main text, Eq.~\eqref{eq: NetworkAction}, to an accuracy of $\eta = 1$. An analysis of the weight distribution of the hidden layer after training shows that for distinguishing between MBL and ETH, the trained network only takes note of the first few input neurons, which correspond to the lowest entanglement eigenvalues. In this case, training has dynamically rediscovered a well known criterion for distinguishing an almost perfectly mixed state (such as a typical ETH state) from a nearly pure state (such as a typical MBL state): the ``Schmidt gap" described in Sec.~\ref{sec: MBLEntEnt}.

\pagebreak 
For input data of kind (b), we find that the network described by Eq.~\eqref{eq: NetworkAction} does not converge to $\eta = 1$ in any reasonable number of training steps. However, when increasing its complexity by introducing a second hidden layer of neurons, we can again easily achieve $\eta = 1$.

In both cases, (a) and (b), the phase diagram is in quantitative agreement with Fig.~\ref{fig: phasediagram}, as long as both regularization and confidence optimization terms are taken into account in the cost function~\eqref{eq: costfunction}. We note that when we drop either of these additions to the cross entropy, the phase diagram obtained after training the respective network to distinguish MBL and ETH with $\eta = 1$ becomes dependent on the type of input data. This indicates that the naive choice of cost function is not restrictive enough to uniquely determine the neural network parameters. Instead it has a manifold of minima which yield inequivalent phase diagrams. This observation reaffirms our choice of cost function, and in particular motivates the extra criterion of confidence optimization we added to encourage a high-confidence classification of transition states, i.e., the second term in Eq.~\eqref{eq: costfunction}.

\section{Hyperparameters and network details}
\label{sec: networkdetails)}
All neural networks used in this work were implemented in Python using Google's TensorFlow\cite{tensorflow2015-whitepaper}. In the main text, it was noted that a network with a large number of hidden layer units needs to be regularized appropriately in order to avoid overfitting. The exact number of hidden layer units is, however, somewhat arbitrary as long as it is large enough with respect to the number of input neurons. For all system sizes considered, with entanglement spectra inputs at most of size $n_1 = {N_A \choose \lfloor N_A/2 \rfloor}|_{N_A=9} = 126$, corresponding to a half-cut of the $N=18$ chain, we chose $n_2=1024$ units for the single hidden layer of our network.

During training, we used an empirically determined step size of $10^{-4}$ for gradient descent on weights initialized around $0$ with a standard deviation of $0.1$. Furthermore, we chose the regularization and confidence optimization parameters $\mu=1$ and $\delta=1$ in Eq.~\eqref{eq: NetworkAction}, respectively. 
We show in Fig.~\ref{fig: costcomparison} that if either confidence optimization or weight decay is ignored, the classification outcome is not consistent with the results from other methods (see also Fig.~\ref{fig: lvlstatisticscomparison} in the main text).

For all calculations, we removed the lowest and highest $10\%$ of all eigenvalues for the data set corresponding to the the pure ETH and MBL regimes, which is then subdivided into training and testing set for cross-validation, and train with randomly chosen batches of $100$ elements in each step. We use the same batch size for the entanglement spectra coming from transition region states, which are required by the confidence optimization term we added to the cost function. We use between $3000$ and $4000$ training steps.

\section{Empirical observations on the network performance}
\label{sec: empirical}
In Sec.~\ref{sec: Results} we noted that the neural network's classification of individual eigenstates for a single disorder realization is not necessarily consistent over multiple training and evaluation runs, even when all hyperparameters and input data are left unchanged. This stems mainly from the random initialization of the weight and bias vectors, not from the randomly chosen training and testing sets. Here we quantify this inconsistency in classification for the case of a single-hidden layer network, given by Eq.~\eqref{eq: NetworkAction}. We trained with entanglement spectra from a single disorder realization of the $N=18$ chain in $3000$ training steps for $50$ times. To distinguish between deviations in confidence and deviations in classification, we define an MBL state as a state that is classified as MBL with a confidence of over $90\%$, and likewise an ETH state as a state which is classified as MBL with a confidence of less than $10\%$ (remember that the confidences for MBL and ETH have to add up to $100\%$ since we use a $\mathrm{Softmax}$ output layer). We then train $50$ times, and consider all possible pairings of the resulting $50$ phase diagrams. It turns out that the fraction of states that change their classification from ETH in one member of a pair to MBL in the other, or vice versa, averages to $0.008$, while the fraction of states that were classified as either MBL or ETH in one member of a pair, but are not confidently classified as either in the other, averages to $0.07$. In Fig.~\ref{fig: training dependence} we present an $\bar{h}$-resolved analysis.

We also compare the network performance when trained on entanglement spectra generated from spin chains of different length, as presented in Fig.~\ref{fig: finitesize}. Here we observe that the transition seems to become sharper for larger system sizes, as expected.\cite{Khemani16} 
However, one has to keep in mind that each panel of Fig.~\ref{fig: finitesize} uses a different disorder realization and hence realization-specific features and those stemming from the change in system size both appear in these plots. 

\bibliography{Deep-learning}

\end{document}